# Hospital Acquired Infections: Advantages of a Computerized Surveillance

**Leonard Mada**

**Formerly Timis branch of the Romanian National Health Insurance; currently in the Department of Biomedical Informatics, Medical University of Timisoara, Romania and Data Management Operations, Romanian branch, Cmed Ltd, Holmwood, UK**

**Abstract**

**Objective:** To assess the advantages of a computerized surveillance system to detect Healthcare-Associated Infections (HAI).

**Methods:** All HAI reported to the Timis County branch of the Romanian National Health Insurance and the Public Health Authority during the year 2007 were collected and assessed for validity.

**Data sources:** International data were used to estimate the true HAI rate. Electronic records from the microbiology lab were used to predict the rate of HAI. Advanced models using DRG data and pharmacy order forms were not possible due to lack of this data.

**Results:** Only 10 out off 17 hospitals in Timis County reported the number of HAI. There were 529 cases during 2007, corresponding to <0.4 cases / 100 hospital admissions compared to 4.5 – 9.1 cases / 100 admissions in the medical literature. We estimated a true number of 6,000 HAI in Timis County during 2007. 29 MRSA and another 88 ceftazidime resistant strains were isolated during July 2007 in one of the hospitals. The total of 117 potential HAI comes close to the 150 HAI predicted for that hospital.

**Conclusions:** Romanian hospitals report less than 10% of the HAI. Available electronic data can be used to estimate the true rate of HAI.

## 1. Introduction

Healthcare-associated infections (HAI) are a major form of morbidity and mortality worldwide. An estimated number of 1.7 million infections occur each year in the USA, with 99,000 deaths directly attributed to the HAI and costs mounting up to $5 – 50 billions per year. [1]

HAI are a major public concern and an important marker of health care quality. The Centers for Disease Control and Prevention (CDC) started to monitor these infections in the USA since 1970, initially through the National Nosocomial Infections Surveillance (NNIS) system, and, since 2006 though the National Healthcare Safety Network (NHSN). [2,3]

The NHSN is a web-based system established in 2005 by the CDC, the Division of Healthcare Quality Promotion, to monitor healthcare-associated adverse events. To this date, over 300 hospitals from the USA were enrolled in the NHSN. [4] Furthermore, 21 US states have adopted legislation mandating reporting of HAI. [5, 6, 7]

Specific legislation does exist in Romania, that mandates hospitals to report HAI to state agencies, both to the county branches of the National Health Insurance, and to the Public Health Authority.

The objective of this study was to analyse the reported HAI rates, to estimate the true incidence of HAI in Romanian hospitals and to test electronic data sources for suitable data to automatically predict the HAI rates.

## 2. Methods

This study was performed while the author was working in the Department of Planing and Forecasting of the Romanian National Health Insurance, the Timis county branch.

HAI data was collected for all hospitals reporting these infections to the Timis County branch of the Romanian National Health Insurance (CJAS Timis). There were 17 hospitals with inpatient wards in the Timis county during 2007. Only 8 of the hospitals did report HAI rates to CJAS Timis, and an additional 2 hospitals did report HAI rates to the Public Health Authoirty (ASP Timis), the remaining hospitals either did not or reported occasionally a zero rate of HAI. All HAI reported through the period January – December 2007 were collected and included in the analysis.

Due to the unrealistic rates of HAI reported to CJAS Timis, hospital managers and laboratory personnel were invited to a training session during April 2007. Furthermore, the importance of HAI surveillance and reporting and the Romanian legislation mandating such a reporting were emphasised. Specific recommendations were issued and a spreadsheet loosely based on the data collected through the US-based NNIS and the NHSN was distributed to all participating hospitals.

The spreadsheet collected both epidemiological patient data, as well as data from microbiology laboratories, including resistance patterns. Specific fields were built in the spreadsheet to ease identification of new HAI cases based on clinical symptoms, microbiological lab data or antibiotic therapy. Any patient presenting with any of the previous 3 features was a candidate case for HAI reporting.

To assess the validity of the reported HAI rates, international rates extracted from the published medical literature were used for a direct comparison. Due to the major discrepancies observed, more realistic estimates were sought. One of the hospitals reported significantly higher rates of HAI, though the number was still lower than the estimate based on international figures. An audit was performed by CJAS Timis on this hospital and the lab data for July 2007 was collected for further analysis.

Electronic DRG data, patient diagnosis and pharmacy order forms were not available, therefore rates of HAI were not feasible to estimate based on reported diagnoses and antibiotic consumption. The author used the microbiological lab data to get a more accurate picture of HAI in this hospital.

The lab data was available in a Comma-Separated-Value (CSV) format, and was poorly structured, with strings of text detailing the microorganism and the resistance patterns in different records. The resistance patterns were entered by hand and contained on three different lines: one composite string with all antibiotics that were resistant for an isolate, one string with antibiotics that were susceptible and one string for intermediate resistance.

Resistance information was extracted from this csv file using a custom written gawk script and various spreadsheet formulas. Methicillin-resistant *Staphylococcus aureus* (MRSA) was identified by searching for the following strings: MRSA, FOX, fox and oxa. The latter searches were performed to detect cefoxitin (FOX, fox) and oxacillin (oxa). Any resistance to either cefoxitin or oxacillin in a *Staphylococcus aureus* strain was interpreted as an MRSA. All matches were manually validated.

The author also counted automatically the number of ceftazidime (CAZ) resistant isolates as a marker of Extended Spectrum Beta-Lactamase production (ESBL). MRSA and CAZ-resistant strains were interpreted as markers of HAI.

Anale. Seria Informatică. Vol. VI fasc. I - 2008
Annals. Computer Science Series. 6th Tome 1st Fasc. - 2008## 3. Results

A number of 17 hospitals were accredited during 2007 in the Timis County. Only 8 of these hospitals reported any HAI to CJAS Timis, while one hospital reported a zero rate during January-March 2007, but never reported again. Two of the hospitals reported HAI only to ASP Timis and this data was requested officially in order to augment the locally available data. The data reported both to CJAS Timis and ASP Timis was cross validated. Five hospitals never reported any HAI rates.

The hospital characteristics include 2 major teaching hospitals with more than 1,000 beds and important intensive care units (ICU) and an additional 4 smaller hospitals with an ICU. The remaining hospitals were primarily medical, psychiatric, terminal illness patient care or rehabilitation hospitals lacking an ICU.

**Table 1.** Healtchcare-associated infections in Timis county reported during 2007. Hospitals 12-17 never reproted any HAI. The data for the first 3 months in 2007 was collected prior to the training session and was added together. Hospitals 1 and 2 reported only to ASP Timis.

| Sp. | ICU* | 3 luni | 04/ 2007 | 05/ 2007 | 06/ 2007 | 07/ 2007 | 08/ 2007 | 09/ 2007 | 10/ 2007 | 11/ 2007 | 12/ 2007 | Total |
|---|---|---|---|---|---|---|---|---|---|---|---|---|
| 1 | Y | 54 | 5 | 19 | 39 | 33 | 57 | 52 | 72 | 66 | 63 | 460 |
| 2 | Y | | | | 1 | 1 | | | 2 | | | 4 |
| 3 | Y | 2 | 11 | | 9 | | | | | | | 22 |
| 4 | N | 1 | 1 | 0 | 4 | 0 | | | 0 | 0 | 2 | 8 |
| 5 | Y | 2 | 2 | 0 | 2 | 0 | 0 | 0 | 0 | | | 6 |
| 6 | Y | 1 | 0 | 2 | 3 | 3 | | | | | | 9 |
| 7 | N | 0 | 0 | 2 | 1 | 2 | 1 | 1 | 1 | 1 | | 9 |
| 8 | Y | 2 | 2 | 1 | 1 | 1 | 0 | 1 | 1 | | | 9 |
| 9 | N | 0 | 0 | 0 | 0 | | | | | | | 0 |
| 10 | N | | 1 | 1 | | | | | | | | 2 |
| 11 | N | 0 | | | | | | | | | | 0 |
| | | 62 | 22 | 25 | 60 | 40 | 58 | 54 | 76 | 67 | 65 | 529 |

* Y = Hospitals with an intensive care unit (ICU); N = without an ICU.

The 10 hospitals included in this study reported a total of 529 HAI during 2007 for an average of 44 HAI per month (see Table 1). The rate was less than 0.4 per 100 patient admissions for 2007 (see Table 2).

Most infections were reported by a single hospital (460/529 HAI, 87%), although only one third of patients are admitted yearly to this

138

hospital. The remaining two thirds of inpatients are treated in the remaining hospitals.

In order to validate the reported results, the author extracted HAI rates from the international literature and compared the HAI rate with the theoretical rate based on those figures. The HAI rate varies between 4.5 – 9.1 cases / 100 admissions in various international studies. [1, 8-11]

**Table 2.** HAI rate in Timis County, Romania compared to data from various international studies.

| HAI per 100 admissions | | | |
|---|---|---|---|
| Country | Rate | Year | Ref |
| USA | 4,5 | (2002) | [1] |
| Greece | 9,1 | (1999) | [8] |
| Denmark | 8,0 | (1999) | [9] |
| Spain | 7,0 | (1997) | [10] |
| Norway | 5,1 | (2002) | [11] |
| **Timis/Romania** | **< 0,4** | **(2007)** | |

A theoretical number of 6,000 HAI occurring yearly in Timis County are estimated based on the international figures for HAI. This is in stark contrast to the actual number of reported cases (529 HAI).

Microbiological data collected for Hospital 1 during an audit session was therefore analysed to get a more accurate assessment of the true HAI rate. Data was available for the month of July 2007.

A number of 431 patients were identified with a positive microbiological culture during this period: 29 were MRSA isolates and 88 were resistant to ceftazidime and therefore potentially ESBL producers. A significant proportion of organisms presented various other resistance patterns, including 160 isolates from 389 tested (41%) resistant to gentamicin and 125 from 346 tested (36%) resistant to ciprofloxacin, although it was not possible to infer any HAI rates from this data because of lacking community resistance patterns.

## 4. Discussion

The aim of this study was to analyse the rate of HAI reported from Romanian hospitals. The number of reported infections was unrealistically small, especially considering also the lack of hygiene and preventive measures in these hospitals.

The medical stuff was advised during a training session in April 2007 to actively perform HAI case detection, while hospital managers were informed of the applicable Romanian legislation and the relevance of accurate HAI reporting. However, the reports continued to provide unrealistic estimates of HAI. Furthermore, neither the medical professionals, nor the hospital managers did acknowledge the real problems posed by HAI.

The author conducted therefore a broader analysis to identify suitable data for automatically identifying HAI, therefore bypassing medical reports. Unfortunately, hospitals do not directly report patient diagnosis and DRG case mix to the Romanian Health Insurance. This data, while best suited to identify HAI, was not available for processing.

Identifying HAI is based both on identifying the actual patients, as well as monitoring infections and resistance patterns in the microbiological lab. Antibiotic consumption based on pharmacy order forms is the third column in a successful HAI surveillance system.

Only microbiological lab data was available to assess the actual HAI rates and this data was available only from one hospital. However, after analysing the microbiological reports, the author clearly identified 431 patients with a clinical isolate. This is especially interesting, because the aforementioned hospital did not possess an infectious diseases ward, all infections being primarily admitted to another hospital. A large proportion of these patients are therefore potential HAI. As a comparison, the hospital reported only 33 cases of HAI during the month of July 2007.

To further strengthen this assumption, the author analysed different resistance patterns in these clinical isolates. MRSA and ceftazidime-resistant isolates were picked up, because they clearly point to a healtchcare-related infection. Community-acquired MRSA was only recently reported in the USA, similar reports lacking from Europe and Romania. Still, most cases occur in health care facilities.

Similarly, ceftazidime (CAZ) is a parenteral $3^{rd}$ generation cephalosporin and is unlikely to be used in the primary health care in Romania, both because of its cost and its parenteral administration. It is a sensible indicator of ESBL production, and therefore it is a potential indicator of HAI.

Other resistance patterns were identified, too. Unfortunately, drugs like fluoroquinolones and aminoglycosides were widely used previously, both in hospitals and the community, making any estimates unreliable.

The author believes that the estimate of 150 HAI for the month of July for hospital 1 is a more realistic one than the 33 cases reported. Unfortunately, other hospitals performed even worse than hospital 1, rarely reporting any HAI. While the HAI rate is arguably negligible in smaller hospitals without surgical wards and an ICU, a total of 6 hospitals do have an ICU, with 2 of these hospitals being big teaching hospitals. The estimated monthly rate of HAI for all hospitals combined is 500 new cases, leaving some 460 cases unreported per month.

The relevance of HAI is related both to the morbidity and mortality and the costs inferred by these infections. International studies showed a direct correlation between infection rates in an ICU and the mortality in that specific ICU. A mortality as low as 10% was reported in Switzerland and as high as 40% was reported in Spain in the SOAP study performed in 198 ICUs throughout Europe, directly correlating with the rate of sepsis and therefore also with the rate of nosocomial infections in the particular hospital. [12]

It is therefore imperative to actively prevent such infections. However, in order to achieve this goal, one has to first acknowledge the existence of these infections and to perform active surveillance to identify the HAI and the cause for that particular HAI case.

Legislation was enacted in a number of states throughout the USA and a number of tools were devised to facilitate this task. [4, 7] In the absence of such resources, electronic medical data should be used to automatically detect HAI. Electronic medical data are widely available in hospitals and should become even more prevalent in the future.

Unfortunately, there are sever issues with the current implementation of these electronic medical systems in Romania, hampering the automatic detection of HAI.

**Conclusions**

Romanian hospitals report only a small proportion of the actual HAI. Electronic medical data has the potential to identify more accurately new cases of HAI, although DRG-data, medical diagnoses and pharmacy order forms were not available for this particular study.

Microbiological electronic lab data was poorly structured and difficult to analyse. Existing medical information systems need to improve in this respect and offer specialized modules to track HAI both at patient level and at microbiological lab level. Modules implementing quality management tasks at hospital level are also desperately needed, especially for managing the ICU, surgical wards and microbiological labs. Furthermore, lab reports should be standardised and stored in a format easy to analyse. Still, it was possible to identify a large number of potential HAI even with this poorly structured lab data, further emphasising the importance of electronic medical data.

Further Informations: Part of this data was presented by the author at the National Lecture on Guidelines and Protocols in ICU ("Cursul National de Ghiduri si Protocoale in ATI"). 5th Edition. Timisoara, November 2007.

Disclaimer: The findings and conclusions in this report are those of the author and do not necessarily represent the views of the Romanian National Health Insurance.

Disclosures: The author is working part time at Cmed, a CRO performing data management operations for the pharmaceutical industry. As such, the author is involved in a number of clinical studies sponsored by the pharmaceutical industry. The author is also a medical advisor for Syonic, a company offering medical informatics systems.